\documentclass[aps,twocolumn,prd,showpacs]{revtex4}
\usepackage{graphicx}
\usepackage{bm}
\usepackage{dcolumn}
\usepackage{multirow}

\newcommand{\bi}{\it\bf}

\newcommand{\muka}{{\mu{\rm K~arcmin}}}

\newcommand{\nhat}{{\hat{\bf n}}}

\newcommand{\rwp}{w_P^{-1/2}}
\newcommand{\rwpe}{w_{P,eff}^{-1/2}}

\begin{document}

\title{Gravitational lensing as a contaminant of the gravity wave signal
in CMB}

\author{Uro\v s Seljak}
\email{useljak@princeton.edu}

\author{Christopher M. Hirata}
\email{chirata@princeton.edu}

\affiliation{Department of Physics, Jadwin Hall, Princeton University,
Princeton NJ 08544, USA}

\date{\today}

\begin{abstract} 
Gravity waves (GW) in the early universe generate $B$-type polarization in
the cosmic microwave background (CMB), which can be used as a direct way
to measure the energy scale of inflation. Gravitational lensing
contaminates the GW signal by converting the dominant $E$ polarization
into $B$ polarization.  By reconstructing the lensing potential from CMB
itself one can decontaminate the $B$ mode induced by lensing. We present
results of numerical simulations of $B$ mode delensing using quadratic and
iterative maximum-likelihood lensing reconstruction methods as a function
of detector noise and beam.  In our simulations we find the quadratic
method can reduce the lensing $B$ noise power by up to a factor of 7,
close to the no noise limit.  In contrast, the iterative method shows
significant improvements even at the lowest noise levels we tested. We
demonstrate explicitly that with this method at least a factor of 40 noise
power reduction in lensing induced $B$ power is possible, suggesting that
$r=P_h/P_R \sim 10^{-6}$ may be achievable in the absence of sky cuts,
foregrounds, and instrumental systematics.  While we do not find any
fundamental lower limit due to lensing, we find that for high-sensitivity
detectors residual lensing noise dominates over the detector noise.

\end{abstract}
\maketitle

\section{Introduction}

CMB polarization is generated by Thomson scattering of photons off free
electrons. To generate polarization one needs an anisotropic distribution
of photons in the electron rest frame (more specifically, a non-vanishing
quadrupole moment) and scattering that couples this angular anisotropy
to the polarization (Thomson scattering in
this case).  These two conditions are satisfied during recombination, when
most of the polarization signal is generated. After recombination the free
streaming of photons leads to a large quadrupole moment, so if some
fraction of the photons is rescattered when the universe is reionized then
a new polarization contribution will be generated at the angular scale of
horizon at reionization. One often refers to the two contributions as the
recombination and reionization components, respectively.

Thomson scattering generates linear polarization only. This is usually
expressed in terms of Stokes parameters $Q$ and $U$, which are coordinate
dependent. They can be decomposed into coordinate independent $E$ and $B$
type polarizations
\cite{1997ApJ...482....6S,1997PhRvD..55.1830Z,1997PhRvD..55.7368K} with
opposite parities.  To linear order in perturbation theory, primordial
scalar (density) perturbations can only generate $E$ polarization, while
gravitational waves (GWs) can generate both scalar $E$ and pseudoscalar
$B$. If the amplitude of the gravity waves is very small relative to
scalars it cannot be isolated from the temperature anisotropies or $E$
polarization due to cosmic variance.  The $B$ polarization is however
insensitive to cosmic variance from scalar modes and is limited only by
instrument noise, foregrounds and sky coverage. This fact generated
attention as a potentially promising tool to detect gravity waves and test
inflation \cite{1997PhRvL..78.2054S,1997PhRvL..78.2058K}. The amplitude of
gravity waves produced during inflation depends on its energy scale:
higher energy scales give larger amplitude of gravity waves.  In terms of
the tensor to scalar power spectrum ratio one has $r=P_h/P_R \propto
V_*^4$, where $P_h$ is the tensor power spectrum, $P_R$ is the scalar
curvature power spectrum and $V_*^4$ is the energy density during
inflation when the present Hubble scale exited the horizon.  $V_*$ has
units of energy and has been termed the ``energy scale'' of inflation.  
We do not know this energy scale, but one of the possibilities is the
scale of grand unification theories (GUTs) at $V_* \sim 10^{16}$~GeV. At
this energy scale the gravity wave contribution is sufficiently large to
be detectable. In this paper we define the tensor to scalar ratio in terms of
their primordial power spectra, rather than the quadrupole moments as
often defined. This has the advantage of relating the tensor to scalar
ratio directly to the inflationary predictions independent of the
cosmological parameters, which affect the CMB anisotropy spectra. For
typical parameters we find $C_2^T/C_2^S\sim r/2$.

A future satellite mission dedicated to $B$ type polarization has been
identified as one of the NASA Einstein probes to be built over the next
decade. One of the outstanding questions regarding such a mission is what
are the required angular resolution and sensitivity to maximize the
science output and at what level do systematics swamp the improvements in
these. It has been pointed out \cite{1998PhRvD..58b3003Z} that
gravitational lensing leads to a generation of $B$ polarization even if
none was present in the early universe. This could limit the extraction of
gravity wave signal if unaccounted for
\cite{2002PhRvD..65b3003H,2002PhRvD..65b3505L}. One can try to reconstruct
the gravitational lensing potential using the non-Gaussian information
present in the CMB data to improve the limits (the large-scale lensing
$B$-modes exhibit higher-order correlations with small-scale polarization
whereas inflationary GW $B$-modes do not). Recent work applied quadratic
estimators \cite{2002ApJ...574..566H} to argue that using these estimators
leads to an order of magnitude improvement for the no noise experiment
\cite{2002PhRvL..89a1304K,2002PhRvL..89a1303K}.  This work has been
interpreted as providing a fundamental limit to the gravity wave
extraction due to the lensing.  However, it is important to note that
these papers do not rule out the
possibility that better reconstruction methods may be constructed. Indeed,
in a recent work we have shown that better estimators are indeed
possible \cite{2003astro.ph..6354H}. We have demostrated explicitly that
one can improve upon the no noise limit of the quadratic estimator.
Indeed, in the idealized case of no noise, perfect resolution and
lensing by a single scalar deflection potential,
 the lensing reconstruction can be achieved {\it exactly}
and the lensing contamination can be removed completely. It is easy to see
why this is so: lensing displaces photons, so one can write the final $Q$
and $U$ polarization in terms of the initial $Q$ and $U$ and lensing
deflection angle $\bi{d}$.  In the absence of gravity waves (null
hypothesis) initial $B$ vanishes and ignoring lensing rotation
(i.e. taking $\nabla\times\bi{d}=0$) the
deflection angle can be written in terms of a scalar field with one degree
of freedom at each point, $\bi{d}=\bi{\nabla}\Phi$. In this case for $N$
pixels there are 2$N$ observables (two at each point, $Q$ and $U$ or $E$
and $B$), and 2$N$ unknowns, initial $E$ and $\Phi$. The number of
unknowns thus equals the number of equations, so one can solve it exactly
in the absence of noise.  It is of course possible that there are
degenerate modes that cannot be reconstructed; however, it has been shown
that the fraction of modes that are degenerate is small and may even be
zero. (See Appendix B of Ref. \cite{2003astro.ph..6354H} for details.)

The idealized case discussed above is unrealistic, since in the real world
noise always limits the achievable sensitivity (in Ref.
\cite{2003astro.ph..6354H} we found that the lensing rotation is not a
limitation at the sensitivity levels that can be achieved in foreseeable
future).  At the same time, if the detector noise is large gravitational
lensing is not limiting the GW detection anyways. As the
detector noise is lowered our ability to clean the lensing contamination 
improves as well and if the scaling between the detector and residual 
lensing noise is linear then lensing may never be the dominating source
of noise. The relevant
question regarding the lensing contamination is thus not whether it
provides a fundamental limit (which remains an open question), but rather
how much does it degrade the gravity wave sensitivity for a given
instrument noise and angular resolution.  We will phrase this question in
terms of a $B$-mode noise power spectrum: the minimum detectable
tensor-to-scalar power spectrum ratio $r$ that can be observed then scales
linearly with it.

The lensing degradation issue is particularly interesting in the context
of the required noise and angular resolution of a future CMB polarization
satellite.  There are important instrument and cost tradeoffs that need to
be included in the design of such a mission. For example, since the bulk
of the gravity wave signal is at large scales one could devise a high
sensitivity low angular resolution instrument costing significantly less
than the equivalent high angular resolution mission.  However, in this
case one would not be able to reconstruct the lensing potential, making
the lensing contamination more significant.  The goal of this paper is to
provide some guidance to these considerations, obtaining the lensing
degradation factors as a function of detector noise and resolution. We
will use both quadratic method \cite{2002ApJ...574..566H} and the
maximum-likelihood method \cite{2003astro.ph..6354H}, which improves upon
quadratic in the low noise regime.  We will ignore other sources of
contamination such as foregrounds and instrument-related issues, which
should be included in the full consideration of pros and cons of a mission
design.  Unless otherwise specified, we will use the fiducial cosmology of
Ref. \cite{2003astro.ph..6354H}.

\section{Lensing and $B$-modes}

In this section, we briefly review techniques for lensing reconstruction
from CMB polarization data.  We then arrive at the main objective of this
paper: to calculate, via simulations, the remaining ``noise'' level
(including lensing residuals) after the lensing reconstruction has been
performed and used to remove lensing $B$-modes.  We compute $\rwpe$, the
white noise power spectrum of the combined instrumental $B$-mode noise and
lensing residuals that remains after lens cleaning.

\subsection{Lensing reconstruction}

While unique features of 
weak lensing effect on CMB power spectrum, such as smoothing 
of the peaks and transfer of power to small scales \cite{1996ApJ...463....1S},
can be used to deduce its presence, it is the nongaussian 
signatures that allow for the lensing reconstruction \cite{1997A&A...324...15B}.
There are several lensing reconstruction methods proposed in the
literature. Simple quadratic estimators \cite{1999PhRvD..59l3507Z} 
have been shown to be
near optimal for the {\slshape Wilkinson Microwave Anisotropy Probe}
({\slshape WMAP}) and possibly {\slshape Planck}
\cite{2000PhRvD..62f3510Z}, but more efficient
quadratic estimators have been shown to improve upon these for low noise,
high angular resolution experiments
\cite{2001ApJ...557L..79H,2002ApJ...574..566H}. Recently we used likelihood
techniques to develop an iterative estimator
\cite{2003PhRvD..67d3001H,2003astro.ph..6354H}, which further improves
upon these if detector noise is sufficiently low, specially for the
polarization data. The corresponding reconstruction errors have been
computed as a function of noise and angular resolution in the absence of
gravity wave contribution \cite{2003astro.ph..6354H}.

We use only polarization information (not temperature) for our lensing
reconstruction in order to reduce the computation time required for the
simulations.  This is obviously a conservative assumption.
But note that -- although temperature anisotropies
are potentially useful for measuring the convergence power spectrum, and
for mapping the convergence on large angular scales ($l\lesssim 200$)
-- most of the lensing-induced $B$-modes come from smaller-scale
convergence modes that are only accessible using polarization data
\cite{2002ApJ...574..566H}.  Additionally, the small-scale temperature
anisotropies are contaminated by secondary processes such as the
kinetic Sunyaev-Zeldovich/Ostriker-Vishniac effect 
and scattering during inhomogeneous reionization,
which have the same spectral dependence as primary CMB fluctuations and
hence can significantly degrade lensing reconstruction from
temperature \cite{2003astro.ph..5471S}.  By comparison, CMB polarization
is expected to be essentially free of secondary scattering contamination
\cite{2000ApJ...529...12H,2003astro.ph..5471S}.

Gravitational lensing re-maps the primary CMB polarization field according
to:
\begin{equation}
\tilde P(\nhat) = P\left( \nhat -
2\nabla\nabla^{-2}\kappa({\nhat}) \right),
\label{eq:p}
\end{equation}
where $P$ is the primary polarization, $\tilde P(\nhat)$ is the lensed
polarization in direction $\nhat$, and
$\kappa$ is the lensing convergence field.
The lensing has two effects on polarization that are of interest
here.  One is the generation of $B$-modes in the lensed polarization
field; the other is the generation of
an anisotropic two-point function:
\begin{equation}
\langle E_{\vec{l}_1} B_{\vec{l}_2} \rangle =
{1\over \pi^{1/2}l^2} C^{EE}_{l_1} \vec{l} \cdot \vec{l}_1
\sin(2\alpha) \; \kappa_{\vec{l}},
\label{eq:ebcorr}
\end{equation}
where $\vec{l}=\vec{l}_1+\vec{l}_2$, $\alpha$ is the angle
between $\vec{l}_1$ and
$\vec{l}_2$, and higher-order terms in $\kappa$ have been
neglected.  (The average is taken over CMB
realizations only, not including noise.)  The $EB$ quadratic estimator
developed by
Ref. \cite{2002ApJ...574..566H} is obtained by taking a
minimum-variance linear combination of $EB$ products subject to the
constraint that the expectation value is
$\kappa_{\vec{l}}$:
\begin{equation}
\hat\kappa^{EB,\rm (quad)}_{\vec{l}} = {A_{EB}(l)\over 2} \sum_{\vec{l}_1}
{ C^{EE}_{l_1}  E_{\vec{l}_1} B_{\vec{l}_2}
\over (\tilde C^{EE}_{l_1} + N^{EE}_{l_1} )( \tilde C^{BB}_{l_2}
+ N^{BB}_{l_2} ) },
\label{eq:eb}
\end{equation}
where $A_{EB}$ is a normalization factor, $N^{EE}_{l_1}$ is the
$E$-mode polarization noise power
spectrum, and $\vec{l}_2 =
\vec{l}-\vec{l}_1$.  Ref. \cite{2002ApJ...574..566H} gives similar
quadratic estimators using the $EE$ products; in this paper we use an
$EE+EB$ quadratic estimator that is a minimum-variance weighting of the
$EE$ and $EB$ products (although almost all of the information for
low-noise experiments comes from $EB$).
For the purpose of cleaning out the lensing $B$-mode, it is
best to Wiener-filter the quadratic estimator
\cite{2002PhRvL..89a1303K} since this minimizes the mean squared
difference between the estimated and ``true''
convergence maps:
\begin{equation}
\hat\kappa^{WF}_{\vec{l}} = { C^{\kappa\kappa}_l \over
C^{\kappa\kappa}_l + \sigma^{\kappa\kappa}_l }
\hat\kappa^{\rm (quad)}_{\vec{l}} ,
\label{eq:kappawf}
\end{equation}
where $\sigma^{\kappa\kappa}_l$ is the power spectrum of the noise in
the estimator
$\hat\kappa^{\rm (quad)}_{\vec{l}}$.  Lens cleaning consists of applying a
de-lensing operation that inverts the mapping of Eq. (\ref{eq:p}).

The quadratic estimator using polarization has excellent performance, in
particular it can ultimately recover the convergence map with
signal-to-noise ratio exceeding unity out to $l\approx 1000$ if the
instrument noise is sufficiently low.  Nevertheless, as is shown in Ref.
\cite{2002ApJ...574..566H}, the accuracy of quadratic lens reconstruction
is fundamentally limited by the $B$-mode cosmic variance.  Put another
way, the estimation of one convergence mode $\kappa_{\vec{l}}$ is
contaminated by $B$-mode power generated by the other covergence modes
$\kappa_{\vec{l}'}$ \cite{2003PhRvD..67l3507K}.

The iterative reconstruction algorithm \cite{2003astro.ph..6354H} avoids
this problem.  While Ref. \cite{2003astro.ph..6354H} introduces the
iterative lensing reconstruction using the likelihood function, the
algorithm used can also be thought of as follows: for low-noise
experiments, the dominant source of uncertainty in the lens reconstruction
is the above-mentioned cross-talk among different convergence modes.  
Once an estimate of the lensing field is available using Eq.
(\ref{eq:kappawf}), the CMB can be de-lensed using the estimated
convergence field.  Since the de-lensed CMB map has most of the
lensing-induced $B$-mode removed, the cross-talk among the convergence
modes is reduced; therefore, re-application of the quadratic estimator
(with different weighting) results in an even lower-noise convergence map.
The reader is referred to Ref. \cite{2003astro.ph..6354H} for
implementation details.  The iterative procedure is only useful if the
instrument noise is below the lensing-induced $B$ signal (5.3 $\muka$ in
the fiducial model at low $l$).  However, as the instrument
noise is reduced, the uncertainty in the iterative lensing reconstruction
is also reduced; indeed, in the absence of foregrounds and the rotational
component of the deflection field, it is not known whether there is {\em
any} fundamental limitation to the reconstruction accuracy that can be
obtained via the iterative estimator.

Since both the quadratic and iterative estimators assume absence of
gravity waves one must be careful when applying them to the case with
gravity waves.  As discussed in the introduction, in the absence of noise
one is simply solving for lensing assuming $B=0$, which is of course
violated if gravity waves are present. The most general approach is to
solve for the tensor $B$ modes, combined scalar and tensor $E$ modes and
lensing potential $\Phi$ simultaneously.  However, since most of the
lensing signal is at high $l$, while most of the intrinsic gravity wave
signal is at low $l$, we can simplify and ignore the information from low
$l$ in the lensing reconstruction, but only use low $l$ in $B$ mode
cleaning.  We have tried cutoffs of $l=50$ and $l=150$ as well as no
cutoff and found no significant difference between them. This is not 
surprising, since number of modes not used in the reconstruction is 
in both cases small compared to the total number of modes. Note that
$l=150$ corresponds roughly to the scale below which the gravity waves
become negligible, while the reionization peak only contributes for $l\ll
50$.

While analytical (Fisher matrix) expressions for noise spectra exist for
both quadratic and iterative methods, they may not be fully
reliable in either case.  The Fisher matrix tends to significantly
underestimate the noise in the iterative method because the
lensing reconstruction error is non-Gaussian and realization-dependent
\cite{2003astro.ph..6354H}. The analytical estimate for the quadratic
estimator error estimate based on the approximation of Gaussian $E$ and
$B$ modes \cite{2002ApJ...574..566H} is more
reliable, but still does not include all terms in the covariance matrix.
Some of these are analytically tractable and have been shown to
increase the covariance by up to 20\%
\cite{2003NewA....8..231C,2003PhRvD..67l3507K}, but other
terms (specifically higher order terms in the deflection angle
$\bi{d}$) have not yet been computed.  The safest
approach is thus to use numerical Monte Carlo simulations, which by
construction include all of the terms.

\subsection{Simulations}

\begin{figure}
\includegraphics[width=3in]{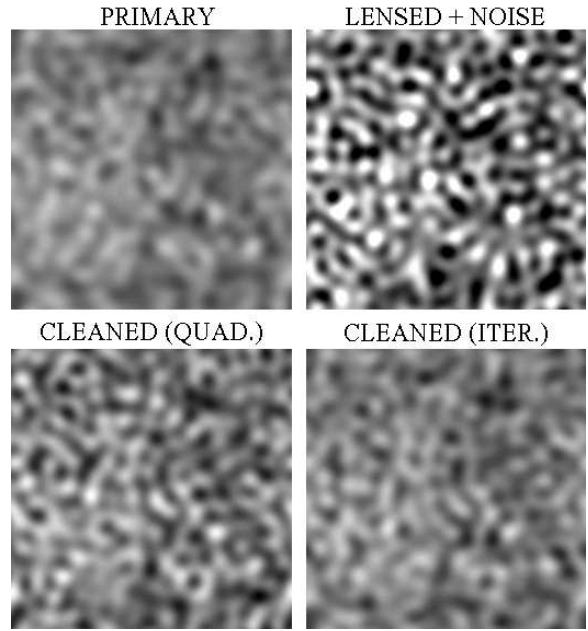}
\caption{
Simulated extraction of a $B$ mode from CMB data
with noise $\rwp = 0.5\mu$K~arcmin and beam FWHM
4~arcmin.  In each panel we have plotted the scalar
quantity $B=\sum_{\bf l} B_{\bf l}e^{i{\bf
l}\cdot\hat{\bf n}}$.  For clarity, only $l<150$
modes are shown.  The widths of the frames are 34 degrees, and the
temperature scale runs from $-0.136\mu$K to $+0.136\mu$K.
{\em Upper left}: The primary $B$ mode.
{\em Upper right}: The $B$ mode after
lensing and addition of 0.5~$\mu$K~arcmin noise.
{\em Lower left}: Recovered $B$ mode after cleaning
with the quadratic estimator.
{\em Lower right}: Recovered $B$ mode after cleaning
with the iterative estimator.
}
\label{fig:simextr}
\end{figure}

We compute the post-cleaning $B$-mode noise power spectrum via
simulation as follows.  First a simulated pure $E$ Gaussian CMB 
polarization field is generated using the unlensed $C^{EE}_l$ power
spectrum from {\sc 
cmbfast} \cite{1996ApJ...469..437S} for the fiducial $\Lambda$CDM
cosmology of Ref. \cite{2003astro.ph..6354H}.  Then a
Gaussian convergence field $\kappa$ is generated, and the polarization $Q$
and $U$ fields are re-mapped according to Eq. (\ref{eq:p}).
We next add noise to the polarization field with power spectrum:
\begin{equation}
C^{EE}_l({\rm noise}) = C^{BB}_l({\rm noise}) =
w_P^{-1} \exp \frac{l(l+1)\theta_{FWHM}^2}{8\ln 2},
\end{equation}
where $\rwp$ is the instrument noise (typically measured in $\muka$) and
$\theta_{FWHM}$ is the full width at half maximum of the instrument's 
beam. The quadratic and iterative lens reconstruction
algorithms are then applied as described in
Ref. \cite{2003astro.ph..6354H}, with the modification that we remove
$l<150$ modes (technically we have set all rows and columns
corresponding to $l<150$ modes to zero in the $\sigma$-matrices of
Ref. \cite{2003astro.ph..6354H}).  Once the estimated convergence field 
$\hat\kappa$ has been determined, it is used to ``de-lens'' the CMB
polarization map, thereby yielding a map of the primary polarization
field. We then compute the $B$-mode power by
averaging $(B_{\bf l}^{\rm (res)})^2$ over the $l<150$ modes in the
simulation.  Each simulation is run on a $2048\times 2048$ square grid
with periodic boundary conditions and grid spacing of 1 arcmin,
corresponding to a total area of 1165 square degrees.  The residual
$B$-mode power spectra quoted here are determined by averaging over
4 such simulations.  We define the ``effective noise'' by:
\begin{equation}
w_{P,eff}^{-1} = C^{BB}_l({\rm residual});
\end{equation}
this has units of $(\muka)^2$ and represents the white noise from
combined instrument noise and lensing residuals that limits
detectability of the GW signal.
Fig.~\ref{fig:simextr} shows an input $B$ polarization map assuming 
$C_2^T/C_2^S=0.012$ (upper left),
no cleaning map (upper right), quadratic cleaning map (lower right) 
and iterative cleaning map (lower right).

\subsection{Results}

The results for the error power spectrum of such a
reconstruction are shown in Table \ref{tab:bres} for a variety of 
noise and beam levels, for both quadratic and iterative methods.
Since the spectra are close to white noise for $l<150$, we
only show the amplitudes and not the full spectrum.
Note that the noise amplitude shown is the total noise and 
contains both lensing and detector noise
contributions.  Also note that for the 20' beam (and for larger beams),
cleaning can actually make the $B$-mode noise worse because the
de-lensing operation transfers power from high-$l$, unresolved CMB modes
down to low $l$.  In principle this problem can be circumvented by
Wiener-filtering the CMB prior to the de-lensing operation; we have not
implemented this because lens cleaning is not useful for such wide beams
anyway.

\begin{table}
\caption{\label{tab:bres}Residual $B$-mode contamination
$\rwpe$ in $\muka$ as a function of the instrument
noise $\rwp$ and beam
FWHM.}
\begin{tabular}{r|dddddd}
\hline
\hline
Beam & \multicolumn{6}{c}{\mbox{Instrument noise
$\rwp$, $\muka$}} \\
FWHM & 6.00 & 3.00 & 1.41 & 1.00 & 0.50 & 0.25 \\
\hline
\multicolumn{7}{c}{\mbox{Quadratic estimator}} \\
\hline
$20'$ & 8.73 & 7.13 & 6.70 & 6.48 & 5.71 & 4.75 \\
$15'$ & 7.73 & 5.11 & 3.92 & 3.64 & 3.28 & 3.06 \\
$10'$ & 7.49 & 4.79 & 3.53 & 3.22 & 2.88 & 2.68 \\
$ 7'$ & 7.32 & 4.59 & 3.29 & 2.98 & 2.62 & 2.40 \\
$ 4'$ & 7.20 & 4.39 & 3.02 & 2.69 & 2.30 & 2.09 \\
$ 2'$ & 7.11 & 4.26 & 2.86 & 2.53 & 2.15 & 1.99 \\
\hline
\multicolumn{7}{c}{\mbox{Iterative estimator}} \\
\hline
$ 7'$ & 7.31 & 4.45 & 2.87 & 2.42 & 1.80 & 1.45 \\
$ 4'$ & 7.17 & 4.23 & 2.56 & 2.07 & 1.39 & 1.00 \\
$ 2'$ & 7.09 & 4.10 & 2.40 & 1.91 & 1.22 & 0.83 \\
\hline
\hline
\end{tabular}
\end{table}

The results can be divided into high, intermediate and low noise regimes.  
For high detector noise, $\rwp>5\muka$, lensing is a minor contributor to
the total noise. Note that this noise level is still a factor of 100 (in
power) lower than expected {\slshape Planck} polarization noise, so
clearly any discussion of lensing induced noise in $B$ is relevant only
for a post-{\slshape Planck} CMB mission dedicated to polarization.  For
$\rwp=6\muka$ a $10'$ beam results in a total rms noise of $7.5\muka$ and
$7'$ beam in $7.3\muka$ for either method.  Without cleaning the combined
lensing and detector noise would be $8\muka$, so cleaning hardly improves
anything at all. Improvements appear when the detector noise drops below
the lensing noise, which for our model is $5.3\muka$. In the intermediate
range ($2$--$5\muka$) the quadratic estimator is very similar to the
iterative method in terms of the residual noise.  For example, for
$\rwp=3\muka$ the residual noise is $4.5\muka$ for $7'$ beam, a factor of
2 in power greater than the detector noise alone and a factor of 2 lower
than no lens cleaning/large beam case.  Going to a $4'$ beam marginally
improves upon this.

Finally, in the low noise regime the iterative method clearly outperforms
quadratic method. The quadratic method bottoms out roughly at
$\rwpe=2\muka$, which is a factor of 7 improvement over the no-cleaning
lensing noise of $\rwpe=5.3\muka$. This bottoming out of the quadratic
method has lead to the suggestion that lensing noise limit may be
fundamental and cannot be improved upon
\cite{2002PhRvL..89a1303K,2002PhRvL..89a1304K}. However, this conclusion
is only valid for the quadratic method, which is not the optimal method in
the low detector noise regime.  We find that the iterative method always
reduces the overall noise as the detector noise is decreased, at least
over the range tested with our simulations (which should cover the range
of interest for the next generation CMB satellite dedicated to polarization). 
At the lowest noise and smallest beam
($0.25\muka$, $2'$) tested in our simulation the lensing noise is reduced
by more than a factor of 40. Further improvements are likely if the
detector noise is reduced below $0.25\muka$, but the iterative method 
becomes very computationally expensive and we have not explored these very
low detector noise cases here.

\begin{figure}
\includegraphics[width=3in]{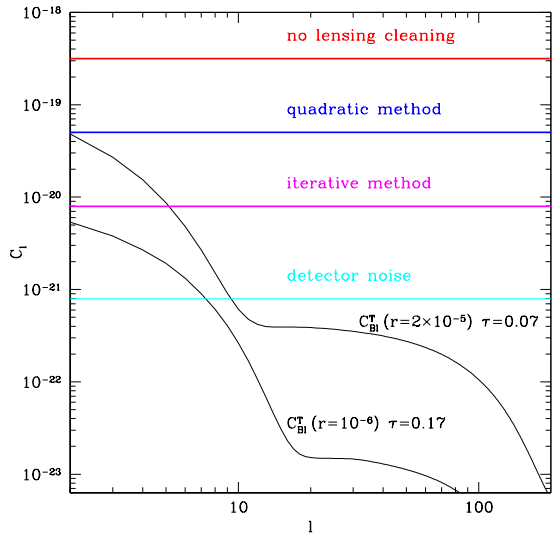}
\caption{Power spectra of noise for $2'$, $0.25\muka$ instrument with
no lensing cleaning, cleaning with quadratic method and cleaning with 
iterative maximum-likelihood method. Also shown are two theoretical 
power spectra for $r=2\times 10^{-5}$ and $r=10^{-6}$. Assuming this instrument 
specifications and iterative method the former can be 
detected (at 2-$\sigma$) 
both in reionization peak ($l<20$) and in recombination peak 
$l>20$), while the latter is detectable for $l<20$ only. 
The noise power spectra have been averaged over the $l<150$ range.
}
\label{fig:pspec}
\end{figure}

For a given noise and lensing induced $B$-mode power spectrum the
resulting uncertainty on $r$ is:
\begin{equation}
\sigma_r^{-2} =
f_{sky}w_{P,eff}^{2} \sum_l {2l+1
\over 2}\left({C_{Tl}^{BB} \over r}\right)^2,
\label{eq:error}
\end{equation}
where $C_{Tl}^{BB}$ is the GW power spectrum of $B$ modes
(Fig.~\ref{fig:pspec}), and $w_P$ is the
inverse noise variance per solid angle per polarization. 
Since $r$ is merely supplying the normalization of the tensor power spectrum,
$C^{BB}_{Tl}/r$ is fixed by the background cosmology. 
We are interested in
large scales only, so the noise spectrum has been approximated as a
constant and taken out of the sum.  It is easy to see from this expression
that the limit on $r=T/S$ is proportional to the noise weight per solid
angle $w_{P,eff}^{-1}$. Therefore, the noise degradation factors are the
same as $r$ degradation factors.  For partial-sky coverage,
Eq.~(\ref{eq:error})
must be modified to take into account sky cuts; while $\sigma_r\propto
f_{sky}^{-1/2}$ for the recombination peak on degree scales, the
reionization peak present at $l<20$ exhibits a much more complicated
dependence on the survey geometry due to cross-leakage of $E$ and $B$
modes induced by, e.g. the Galactic Plane cut \cite{2002PhRvD..65b3505L}.  
Nevertheless, it remains true even in the presence of sky cuts that
$\sigma_r\propto w_{P,eff}^{-1}$ if
only the pure $B$-modes are used for GW searches.
(But note that for the reionization peak, the uncertainty on $r$ can be
non-Gaussian due to the small number of modes, which complicates
hypothesis tests for a GW contribution \cite{2002PhRvD..65b3505L}.)

\section{Discussion and Conclusions}

While the degradation factors can be computed independently of the 
theoretical spectrum, the actual achievable values of $r$ depend on it.
It is worth considering the reionization and recombination peaks
separately.
While the reionization peak gives typically higher signal to noise if
$f_{sky}=1$, it
depends sensitively on the Thomson scattering optical depth $\tau$ due to
reionization, which is still rather uncertain (although this may improve
with future observations of the reionization-induced 
$TE$ correlation, which was recently detected by {\slshape WMAP}
\cite{2003ApJS..148....1B,2003ApJS..148..161K}).
In addition, incomplete sky coverage and foregrounds are particularly
worrisome on large scales, so the reionization peak may be more difficult
to observe than the recombination peak at $l\sim 100$.

As an example, for $l<20$ and optical depth $\tau=0.17$ one finds
$\sigma_r=5\times 10^{-7}$ for $f_{\rm sky}=1$ and $\rwpe=0.8\muka$, which
is the lowest noise level found in our simulations.  The energy scale of
inflation scales as $r^{1/4}$, so in this case the minimum energy scale
one can detect at $3\sigma$ is $V_* \sim 10^{15}$~GeV. For full-sky
coverage, the signal-to-noise scales somewhat less rapidly than $\tau^2$
and reducing the optical depth to $\tau=0.07$ increases $\sigma_r$ by a
factor of 3. Additional reduction will be caused by incomplete sky
coverage, which will cause leakage of $E$ into $B$
\cite{2002PhRvD..65b3505L,2003PhRvD..67b3501B}; this degradation will be
worst for models with late reionization because this pushes the $B$
reionization peak to the low multipoles where sky-cut effects are most
severe.  For the recombination peak at $l>20$ one has $\sigma_r=10^{-5}$
for $\rwpe=0.8\muka$, which is a factor of 20 worse than reionization peak
if $\tau=0.17$ and a factor of 7 worse if $\tau=0.07$. The corresponding
energy scale that can be detected at $3\sigma$ is $V_* \sim 2.3\times
10^{15}$~GeV.

To summarize, in this paper we present a detailed numerical study of how
well can one clean $B$-type polarization of the contamination caused by
gravitational lensing.  We find that quadratic methods are able to reduce
the noise power by up to a factor of 7, while our iterative method is able
to reduce the noise significantly beyond that, at least a factor of 40 in
our simulations.  With this method we do not find any fundamental lower
limit caused by gravitational lensing, in the sense that over the
range of parameters considered, reducing the
instrument noise always leads to a reduction in lensing noise as well.
However, the scaling is sublinear, so at low detector noise levels lensing
noise dominates over instrument noise. With the noise cleaning one can
achieve tensor to scalar ratios as low as $r\sim 10^{-6}$ and possibly 
even lower, which
should allow us to differentiate between different models of structure
formation with high precision.  In particular, inflationary models with an
energy scale significantly below $10^{15}$~GeV, as well as
cyclic/ekpyrotic models
\cite{2001PhRvD..64l3522K,2002Sci...296.1436S,2003hep.th....7170B} predict
that the gravity waves should be negligible and so their predictions
could be falsifiable with the future CMB polarization studies.

\acknowledgments

U.S. is supported by
Packard Foundation, Sloan Foundation,
NASA NAG5-1993 and NSF CAREER-0132953.
C.H. is supported by a fellowship from the NASA Graduate Student
Researchers Program (GSRP).

\bibliography{cosmo,cosmo_preprints}

\end{document}